\documentclass[12pt,letterpaper]{article}  

\usepackage[includeheadfoot,
            marginratio={1:1,2:3}, 
            width=412pt, 
            height=688pt,]{geometry}

\usepackage{amsmath}
\usepackage{amsfonts}
\usepackage{amssymb}
\usepackage{mathtools}
\usepackage{mathalfa}
\usepackage{graphicx,caption,subfigure}
\usepackage[dvipsnames]{xcolor} 
\usepackage{tikz} 
\usetikzlibrary{decorations.pathreplacing,calc}
\usepackage{booktabs}
\usepackage{adjustbox}
\usepackage[utf8]{inputenc}
 \usepackage[splitrule]{footmisc} 
\usepackage{soul}

\usepackage{empheq}
\usepackage{paralist}
\usepackage{cite}
\usepackage[normalem]{ulem}

\usepackage[colorlinks=false,urlbordercolor=red]{hyperref}
\usepackage{tensor}

\newcommand{\nc}{\newcommand}
\nc{\lb}{\llbracket}
\nc{\rb}{\rrbracket}
\nc{\gl}{\llbracket}
\nc{\gr}{\rrbracket}
\nc{\del}{\partial}
\nc{\tri}{\hspace{-3.5pt}\vartriangle\hspace{-3.5pt}}
\nc{\blacktri}{\blacktriangle}

\nc{\eq}[1]{\begin{equation}
                     \begin{split} #1 \end{split}
                     \end{equation}}
\nc{\ov}{\overline}

\nc{\fa}{\hat}
\nc{\fb}{\MakeUppercase}
\nc{\fc}{\tilde }
\nc{\Lie}{{\cal L}} 
\nc{\lambdabar}{{\mkern0.75mu\mathchar '26\mkern -9.75mu\lambda}}

\allowdisplaybreaks[2]
\numberwithin{equation}{section}

\DeclareUnicodeCharacter{2212}{-}
\begin{document}

\vspace*{-1.5cm}
\begin{flushright}
  {\small
  MPP-2022-57\\
  }
\end{flushright}

\vspace{1.5cm}
\begin{center}
  {\Large
    Dynamical Cobordism  of a Domain Wall\\[0.3cm]
   and its Companion Defect 7-brane 
} 
\vspace{0.4cm}

\end{center}

\vspace{0.35cm}
\begin{center}
Ralph Blumenhagen, 
Niccol\`o Cribiori, Christian Knei\ss l and Andriana Makridou
\end{center}

\vspace{0.1cm}
\begin{center} 
\emph{
Max-Planck-Institut f\"ur Physik (Werner-Heisenberg-Institut), \\[.1cm] 
   F\"ohringer Ring 6,  80805 M\"unchen, Germany } 
   \\[0.1cm] 
 \vspace{0.3cm} 
\end{center} 

\vspace{0.5cm}

\begin{abstract}
Starting from an already known solution in the literature, we study the dynamical cobordism induced by the backreaction of a non-supersymmetric, positive tension domain wall in string theory. 
This could e.g.~be a non-BPS D8-brane of type I or a $\overline{D8}/O8$ stack of a non-supersymmetric type IIA orientifold.  
The singularities which typically appear indicate either an inherent inconsistency of this background or the required presence of a suitable defect, as predicted by the cobordism conjecture.
We provide evidence that this end-of-the-world $7$-brane is explicitly described by a new kind of  non-isotropic solution of the dilaton-gravity equations of motion. 
Intriguingly, on the formal level this solution turns out to be  closely related  to the initial solution for the non-supersymmetric domain wall.
 \end{abstract}

\thispagestyle{empty}
\clearpage

\section{Introduction}
\label{sec:intro}

One of the most  fundamental swampland conjectures \cite{Palti:2019pca,vanBeest:2021lhn,Grana:2021zvf} is the absence of global symmetries in quantum gravity \cite{Banks:2010zn,Banks:1988yz}.
The notion of symmetry can be extended to include higher-form global symmetries \cite{Gaiotto:2014kfa}, which should also vanish. 
Thus, taking topological global charges into account, led to the recent proposal that the cobordism group in quantum gravity needs to be trivial \cite{McNamara:2019rup}. 
Starting with a non-trivial cobordism group, the induced global symmetry can either be gauged or broken.
Its breaking requires the introduction of defects, which have to cancel the cobordism charge.
We refer to \cite{GarciaEtxebarria:2020xsr,Montero:2020icj,Dierigl:2020lai,Hamada:2021bbz,Debray:2021vob,McNamara:2021cuo,Blumenhagen:2021nmi,Andriot:2022mri} for more recent developments related to the cobordism conjecture.

In the series of papers \cite{Buratti:2021yia,Buratti:2021fiv,Angius:2022aeq}, it has been argued and supported by many examples that, from the effective supergravity point of view, certain non-trivial cobordism groups are related to dynamical tadpoles or more general scalar potentials.
This is the case in the presence of running solutions of the equations of motion that feature end-of-the-world (ETW) branes at finite space-time distance, where in addition some scalar fields go to infinite distance in field space.
This picture was dubbed Local Dynamical Cobordism, since it is based on the interplay between the dynamics of the scalars and the breaking of cobordism symmetries, while it can be given a local description in large classes of models.
It is the purpose of the current work to continue along these ideas and provide one more non-trivial example. We will give evidence that breaking a certain global cobordism symmetry requires the existence of a new object in the effective theory, which can be detected by a novel kind of non-isotropic solution of the dilaton-gravity equations of motion.

Concretely, we consider the backreaction of a gauge neutral, non-supersymme\-tric 9-dimensional domain wall  carrying only a positive tension and coupling to the dilaton like a D-brane, i.e.~with the factor $\exp(-\Phi)$ in its action. 
Such a domain wall arises e.g.~as the non-BPS D8-brane of type I string theory or as a local R-R tadpole-free $16\times\overline{D8}+O8^{++}$ stack in a certain non-supersymmetric orientifold, which is the T-dual of the Sugimoto model \cite{Sugimoto:1999tx}.
In short, we will  address  such an object as a neutral domain wall.

Fortunately, this backreaction problem has already been addressed in 2000 by Anamaria Font and one of us \cite{Blumenhagen:2000dc}.
Similar to the solution for the backreacted original Sugimoto model \cite{Dudas:2000ff}, it was shown in \cite{Blumenhagen:2000dc} that the gravity and dilaton equations of motion of the T-dual setup do not admit a solution with nine-dimensional Poincar\'e symmetry (along the longitudinal directions of the brane), but induce a spontaneous compactification on a transversal circle times a non-trivial longitudinal direction. In \cite{Blumenhagen:2000dc}, two explicit solutions were presented, where for the first the longitudinal direction was non-compact, whereas for the second it was a finite size interval.

In this work, we revisit these solutions in light of the recent developments around (dynamical) cobordism.
It was already realized in \cite{Blumenhagen:2000dc} that in such
solutions there appear additional
singularities, where e.g.~the string coupling diverges. 
At that time this was a rather puzzling issue, but in view of the
improved understanding of cobordism symmetries in quantum gravity,
it might just be pointing to the required existence of an ETW defect
7-brane, curing the singularity (and breaking the symmetry) and making
the whole solution consistent. Let us also mention that the possible presence of 8-branes in the
original
Sugimoto model has been proposed in \cite{Antonelli:2019nar}, although from a different perspective.

In section \ref{sec:nonbps}, we first review the two solutions of \cite{Blumenhagen:2000dc} for the backreaction of such a neutral domain wall and show that the so called solution II$^+$ fits especially well into the recent discussion of Local Dynamical Cobordism \cite{Buratti:2021yia,Buratti:2021fiv,Angius:2022aeq}.
Analyzing  the behavior close to the singularity, we find that it is
of logarithmic type, something one would indeed expect from a
codimension-two object. This is strong indication for the presence
of ETW 7-branes at the endpoints of the finite size interval.

As a new development, in section \ref{sec:etw7} we construct  the (local) solution of the dilaton-gravity equations of motion around the potentially present ETW-7-brane.
Since this object will also just carry tension, we cannot expect to find a familiar solution, preserving both 8D Poincar\'e and 2D rotational symmetry.
Given that we do not want to break the 8D Poincar\'e symmetry (and be probably forced to introduce more defects of lower dimensionality), we make an ansatz for a solution breaking the 2D rotational symmetry.
It turns out that the resulting equations of motion do admit a solution that has a striking similarity to the neutral domain wall we started from.
The ETW 7-brane turns out to have positive tension as well and shows the expected logarithmic singularities. 
The coupling to the dilaton is $\exp{(-2\Phi)}$ in string frame, which does not correspond to any known (to us) codimension-two object in string theory. 
It would thus be interesting to confirm the presence of this candidate new object in string theory with an independent analysis.

\section{Backreacted  Domain Wall}
\label{sec:nonbps}

We consider a  neutral domain wall  configuration in ten dimensions carrying positive tension $T$  and being located at the position $r=0$ in the transversal directions. 
Since it carries no other gauge charge, at leading order its supergravity
action assumes the form
\eq{
        S={\frac{1}{2 \kappa_{10}^2}}\int d^{10}x  \sqrt{-G} \left( \mathcal{R}-{\frac 12} (\partial \Phi)^2 \right) - T \int d^{10}x \sqrt{-g}\, e^{{\frac 54} \Phi}\, \delta(r),
      }
where $G_{MN}$ denotes the metric in ten dimensions and $g_{\mu\nu}$ the metric on the nine-dimensional worldvolume of the brane. Moreover, one has $\kappa_{10}^2=\ell_s^8/(4\pi)$, with $l_s$ the string length.
Note that this is actually the same action as considered in \cite{Blumenhagen:2000dc}.
The only slight difference is the presence of a single source instead of two. Indeed, the second source would sit at $r=R$ and it is here taken into account by extending periodically the solution beyond $r \in [-\frac R2, \frac R2]$.

The resulting  gravity equation of motion reads
\eq{
  \label{Einsteineom}
                  \mathcal{R}_{MN} -{\frac 12} G_{MN} \mathcal{R} -{\frac 12}&
                  \left(\partial_M\Phi \partial_N \Phi -{\frac 12} G_{MN}
                    (\partial \Phi)^2\right)=\\
                  &
                  \hspace{4cm}
                  -\lambda \,\delta_M^\mu \delta_N^\nu \,g_{\mu\nu} \,\sqrt{\frac gG}\, 
                e^{{\frac 54} \Phi} \,\delta(r),
                }
with $\lambda=\kappa_{10}^2 T$. In addition, there is only the  dilaton equation of motion 
\eq{
  \label{dilatoneom}
              \partial_M \left( \sqrt{-G}\, G^{MN} \,\partial_N
                \Phi\right)={\frac 52}\lambda \sqrt{-g} \, e^{{\frac 54} \Phi} \,\delta(r)\,.
}
These are non-linear equations that share some features with the
backreaction of BPS-branes but, since  they do not involve any
$p$-form gauge field, they  are actually of non-supersymmetric type
and in general much harder to solve.

\subsection{Two solutions breaking 9D Poincar\'e symmetry}

These equations do not admit a solution preserving 9D Poincar\'e
invariance. 
However, in \cite{Blumenhagen:2000dc} a solution was found that preserved 8D Poincar\'e invariance featuring a single  non-trivial longitudinal direction $y$.
The general ansatz for the metric was 
\eq{
  \label{metricansatz}
  ds^2=e^{2{\cal A}(r,y)} ds_8^2 + e^{2{\cal B}(r,y)}( dr^2 +dy^2),
}
with a separated dependence of the warp factors $\mathcal{A}$, $\mathcal{B}$ and the dilaton $\Phi$ on the coordinates $r$
and $y$, i.e. 
\eq{
       &{\cal A}(r,y)=A(r) + U(y)\,,\quad {\cal B}(r,y)=B(r) +
       V(y)\,,\\
       &\Phi(r,y)=\chi(r) + \psi(y)\,.
}
For such a separation of variables, by redefining $r$ and $y$ the
ansatz \eqref{metricansatz} is actually the most general one.

The equations of motion lead to five a priori independent equations.
The one related to the variation  $\delta G^{\mu\nu}$ is
\eq{
  \label{eommunu}
  \Big( 7 A'' +28 (A')^2 +B'' +{\textstyle \frac 14}(\chi')^2\Big)
  +&\Big( 7 \ddot U + 28 (\dot U)^2+ \ddot V + {\textstyle \frac 14} (\dot
  \psi)^2\Big)\\[0.1cm]
    &
    \hspace{3cm}
    =-\lambda\, e^{B+V}\, e^{{\frac 54} \Phi} \, \delta(r)\,.
  }
 The prime denotes the derivative with respect to $r$ and the dot the derivative with respect to $y$.
 For the two variations $\delta G^{rr}$ and $\delta G^{yy}$, we obtain
\eq{
  \label{eomyyrr}
   &\Big( 28 (A')^2 +8A' B' - {\textstyle{\frac 14}}(\chi')^2\Big) +
  \Big( 8 \ddot U+ 36 (\dot U)^2- 8 \dot U\dot V + {\textstyle{\frac 14}} (\dot
    \psi)^2\Big)=0,\\[0.2cm]
  &\Big( 8 A'' +36 (A')^2 -8A' B' +{\textstyle \frac 14}(\chi')^2\Big)+
  \Big(  28 (\dot U)^2+ 8 \dot U\dot V - {\textstyle \frac 14} (\dot
    \psi)^2\Big)\\[0.1cm]
  &
  \hspace{6cm}
  =-\lambda\, e^{B+V}\, e^{{\frac 54} \Phi} \,
  \delta(r)\,,\\
}
and for  the off-diagonal $\delta G^{ry}$
\eq{
  \label{eomyr}
  &-8 A' \dot U  +8 B' \dot U +8 A' \dot V -{\textstyle \frac 12} \chi' \dot\psi=0  \,.
}
Finally, the dilaton equation of motion becomes
\eq{
  \label{eomdila}
     \Big(\chi''+8A'\chi'\Big)+\Big(\ddot\psi +8 \dot U \dot
       \psi\Big)={\frac 52}\lambda\, e^{B+V}\, e^{{\frac 54} \Phi} \, \delta(r)\,.
}
Note that taking the sum of the two equations in \eqref{eomyyrr} gives the simpler equation
\eq{
  \label{eomsum}
        8\Big( A'' +8 (A')^2\Big)+8\Big(\ddot U +8 (\dot U)^2\Big)=-\lambda\, e^{B+V}\, e^{{\frac 54} \Phi} \,  \delta(r)\,.
}

One first solves these equations in the bulk and then implements the $\delta$-source via a jump of the first derivatives $A', B', \chi'$ at $r=0$. 
We proceed completely analogously to \cite{Blumenhagen:2000dc}, so
that we can keep the presentation short and essentially just describe
the two solutions found there.\footnote{In fact, in this new approach we did not find any further relevant solution, either. In appendix \ref{sec_append} we present a physically less interesting solution we have identified.}

\subsubsection*{Solution I}

From \eqref{eomsum} one learns that both expressions in the brackets
need to be constant (away from the sources). 
This forces one of the two functions $A(r)$ or $U(y)$ to be of trigonometric type and the other one in general of hyperbolic type. 
It turns out that all equations of motion and boundary conditions are satisfied by the following functions 
\eq{
\label{SolIexpl}  
        A(r)=B(r)&={\frac 18} \log\Big\vert \sin\left[ {\textstyle 8K (|r|
          -{\textstyle {\frac R2}})}\right] \Big\vert,\\[0.1cm]
        \chi(r)&=-{\frac 32} \log\Big\vert \tan\left[ 4K (|r|
          -{\textstyle {\frac R2}})\right] \Big\vert + \phi_0,\\[0.1cm]
        U(y)&=-K\, y,\,\qquad \psi(y)=V(y)=0\, ,
}
where the integration constant $\phi_0$ is related to the string coupling constant.
Moreover, the parameters appearing have to satisfy 
\eq{
  \cos(4KR)={\frac 35}\,,  \qquad e^{{\frac 54}\phi_0}= 3 \left(\frac 52\right)^{\frac 18} \frac{K}{\lambda}\sim \frac{1}{\lambda R}\,.
}
The second  relation nicely reflects that the compactness of the $r$ direction is a consequence  of the backreaction of the  neutral domain wall.
Indeed, sending the string coupling to zero, the compact space decompactifies.

Hence, the direction transversal to the  neutral domain wall is
spontaneously compactified on  a topological $S^1$ of a size related
to $R$.\footnote{Another possibility would be to consider the $r$-direction to be an interval of finite proper size proportional to $R$. Since there is no distinguished singular point in the $y$-direction, in this way  one would be led to a pair of  ETW 8-branes at $r=-R/2$ and $r=R/2$. However, an 8-brane is not consistent with the singularities \eqref{solasings} of the solution at these points and is also not expected from the cobordism argument (which suggests an ETW 7-brane). Moreover, since there are really periodic trigonometric functions appearing in the solution \eqref{SolIexpl}, we consider this as a clear indication that the $r$-direction is circular.}
However, as opposed to the upcoming solution II, the proper length of the $y$-direction is infinite so that there are no walls of nothing in this solution (equivalently, walls of nothing are infinitely far away and thus cannot be captured by the given effective description). In other words, we necessarily ask for the presence of a coordinate of finite proper length in order to give an interpretation in terms of local dynamical cobordism.
Moreover, the solution as it stands  is inconsistent in the sense that at $r=\pm R/2$ there appear logarithmic singularities in the warp factors and the dilaton
\eq{
 \label{solasings} 
  A(r)=B(r)\sim {\frac 18}\log\rho\,,\qquad
  \chi(r)\sim-\frac 32 \log \rho,
}
with $\rho=|r|-R/2$.
This leads to curvature singularities, where the string coupling diverges.
The appearance of a singularity at $\rho=0$  was already observed in \cite{Blumenhagen:2000dc}, but in those early days it could not be clarified what it was indicating. 
We will come back to this issue after discussing Solution II.

\subsubsection*{Solution II}

\begin{figure}[ht]
  \centering
  \includegraphics[width=4.7cm]{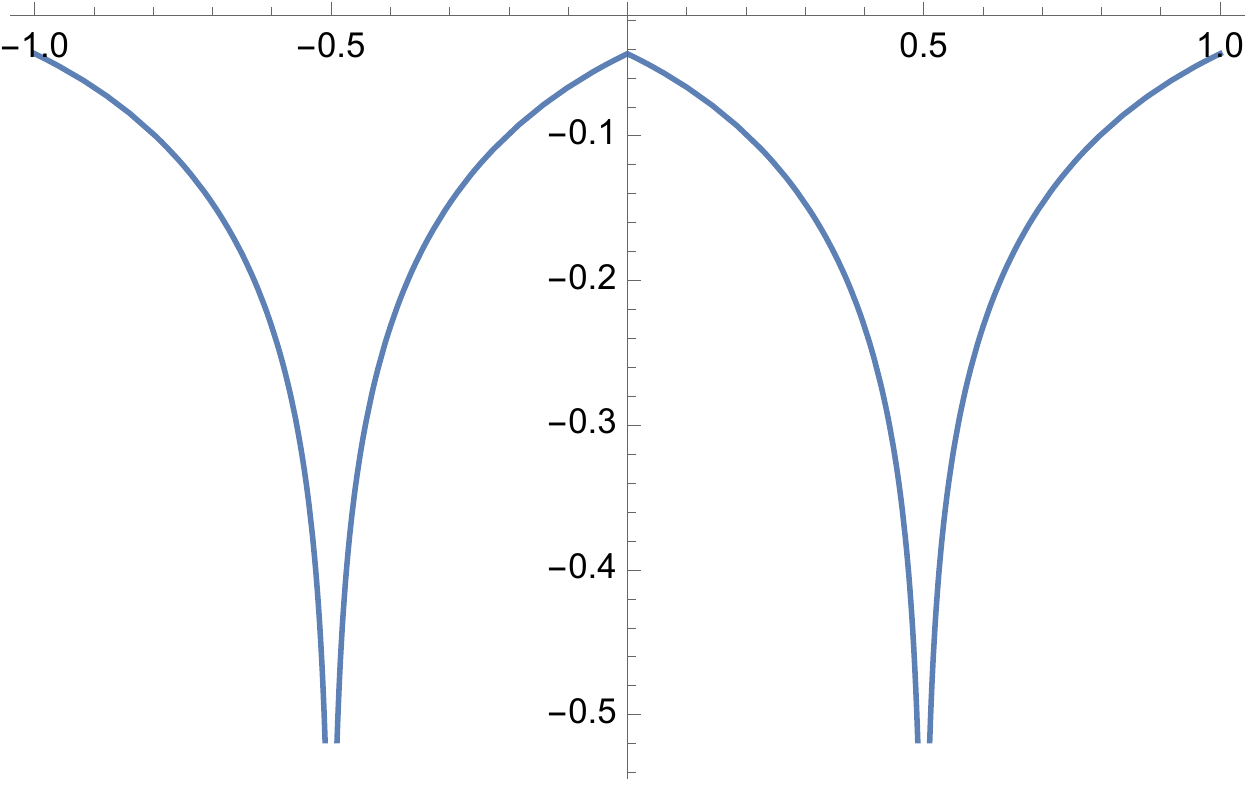}
  \includegraphics[width=4.7cm]{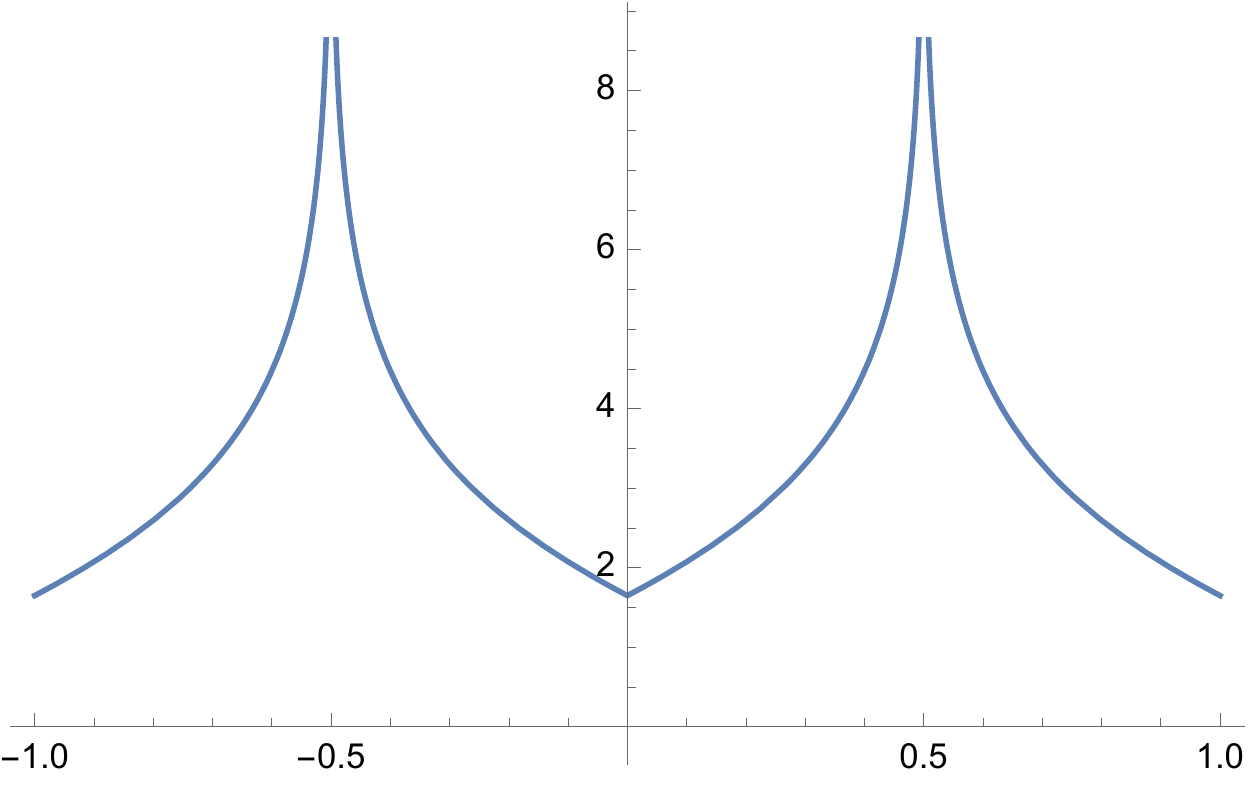}
  \includegraphics[width=4.7cm]{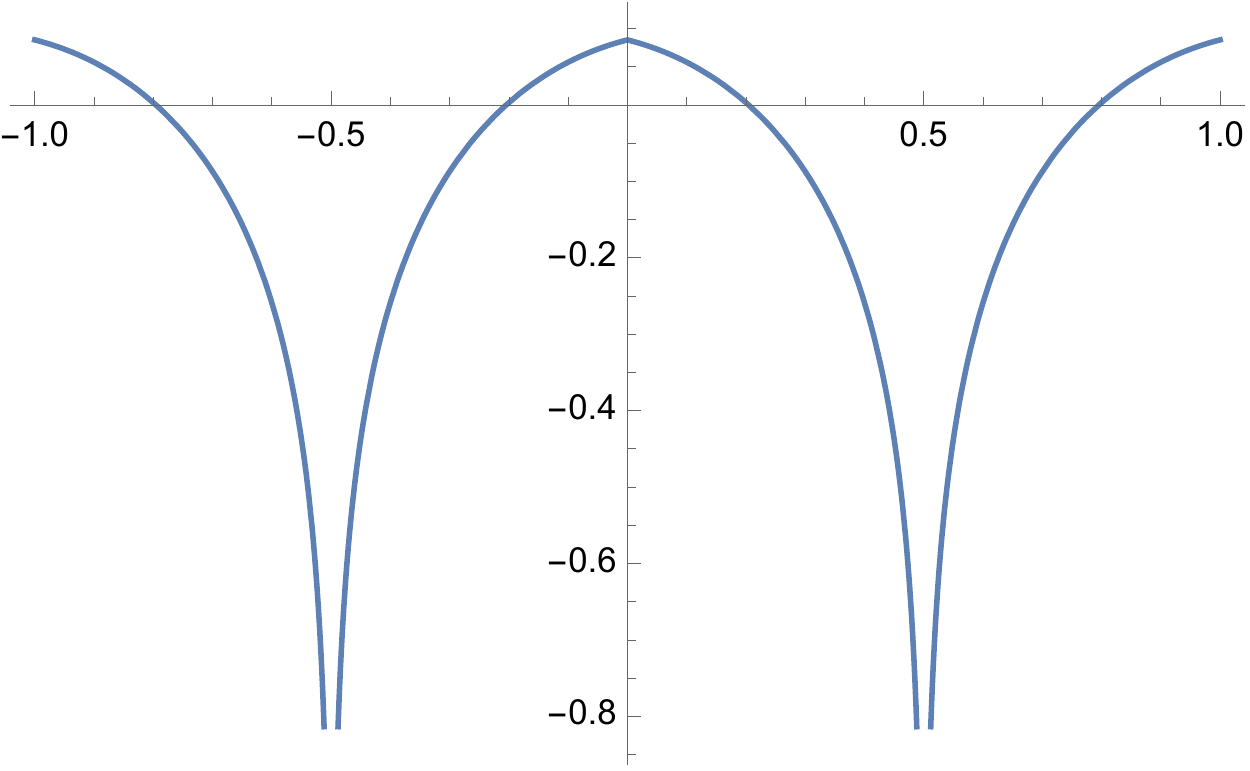}
\begin{picture}(0,0)
    \end{picture}
    \caption{For Solution II$^+$, the three functions $A(r), \chi(r)$ and $B(r)$ are shown
      for $R=1$ in a double period $r\in [-R,R]$. }
  \label{fig:AchiB}
\end{figure}

Let us now review the more involved Solution II to the equations of motion.  
The equations in the bulk still admit three free parameters, $\alpha, K, R$, which are further restricted by  implementing the 8-brane boundary conditions at $r=0$.
Eventually, it was found that the $r$-dependent solutions satisfying
the proper jump conditions at $r=0$ are 
\eq{
 \label{sol2r} 
        A(r)&={ \frac 18} \log\Big\vert \sin\left[ {\textstyle {8 K}} (|r|
          -{\textstyle {\frac R2}})\right] \Big\vert,\\[0.1cm]
        \chi(r)&={\frac{\alpha^\pm}{8}} \log\Big\vert \sin\left[ {\textstyle {8 K}} (|r|
          -{\textstyle {\frac R2}})\right] \Big\vert \mp 2 \log\Big\vert \tan\left[ {\textstyle {4 K}} (|r|
          -{\textstyle {\frac R2}})\right] \Big\vert + \phi_0,\\[0.1cm]
         B(r)&={\frac{\mu}{8}} \log\Big\vert \sin\left[ {\textstyle {8 K}} (|r|
          -{\textstyle {\frac R2}})\right] \Big\vert \mp {\frac{\alpha^\pm}{8}} \log\Big\vert \tan\left[ {\textstyle {4 K}} (|r|
          -{\textstyle {\frac R2}})\right] \Big\vert ,
      }
with  $r\in[-{\frac R2},{\frac R2}]$.      
The parameters appearing above are defined as
\eq{
  \mu={\frac{\alpha^2}{16}}+{\frac{5\alpha}{4}}+1\,,\qquad {\rm for}\quad
  \alpha^\pm= -4(5\mp 4\sqrt{2})\, ,
}  
where the latter are the positive and the negative  root  of 
\eq{ 
     \alpha^2+40\alpha -112=0\,.
   }
Moreover, consistency of the boundary conditions implies
\eq{
  \cos(4KR)=\pm {\frac{16}{\alpha^\pm+20}}={\frac{1}{\sqrt 2}},
}
with the minimal solution $K=\pi/(16R)$. 
The integration constant $\phi_0$ is related to the string coupling constant and has
to satisfy
\eq{
  \label{dilatonvev}
        e^{{\frac 54}\phi_0}={\frac{\pi}{\lambda R}} e^{-B-{\frac 54}(\chi-\phi_0)}\Big\vert_{r=0}={\frac{\pi}{\lambda R}} \sqrt{2} \left(\sqrt{2}-1\right)^{2\sqrt{2}}\,.
      }
Again, we have a compact direction parametrized by $R$ which becomes non-compact as the string coupling is sent to zero. 
The solutions for the $y$-dependent functions are a bit simpler and read
\eq{
            U(y)&={\frac 18} \log\Big( \cosh\left[ 8K \, y \right] \Big)  
              \,,\quad \psi(y)=\alpha U(y)\,,\quad
        V(y)=-{\frac{5\alpha}{4}} U(y)\,,
 }
i.e.~all three functions are proportional.
Notice also that we have chosen one integration constant such that the solution is symmetric around $y=0$. 
We will denote the present two solutions as II$^\pm$, to indicate that they refer to the roots $\alpha^\pm$ respectively.
As we will see shortly, only Solution II$^+$  leads to a finite size of the $y$-direction. 
For this solution, in figure \ref{fig:AchiB} we display the  three functions depending on $r$, showing their characteristic features.
\vspace{0.2cm}

\noindent
One can see that by construction they all have a kink at the location $r=0$ of the neutral domain wall, but  exhibit a singularity at $r=\pm R/2$.
Introducing the coordinate  close to the singularity, $\rho=r-R/2$, and expanding $\sin[8K(|r|-{\frac R2})]\simeq {8K} \rho$, the behavior of these three functions close to $\rho=0$ is 
\eq{
  \label{singula}
  A(\rho)\simeq {\frac 18}\log\rho\,,\quad
    \chi(\rho)\simeq {\frac 18}(\alpha^+-16)\log\rho\,,\quad
      B(\rho)\simeq {\frac 18}(\mu-\alpha^+)\log\rho\,.
}
Since $(\alpha^+-16)<0$ while $(\mu-\alpha^+)>0$, for $\rho \to 0$ the warp factors $A$ and $B$ go to zero while the string coupling $g_s=\exp(\Phi)$ goes to infinity, i.e.~we are facing a strong coupling singularity.

\subsection{The geometry of the solution}

In this section, we would like to discuss the stringy geometry of Solution II in more detail. For this reason, we mainly work in the string
frame, while we will mention some results in the Einstein frame as
well.\footnote{We recall that the string frame metric is obtained by
  multiplying the Einstein frame metric \eqref{metricansatz} with $\exp((\Phi-\phi_0)/2)$.}

We start by studying the geometry along the $y$-direction. In string frame the proper length (at fixed $r$) is 
\eq{
         L_y=  \int_{-\infty}^\infty  dy\, e^{V(y)+{\frac 14}\psi(y)}=\int_{-\infty}^\infty  dy\,
           \Big(\cosh\left[ {\textstyle {\frac{\pi}{2R}}} y \right]\Big)^{-{\frac{\alpha}{8}}}\,.
}
By inspection of the integral, one can see that this length can only be finite for positive $\alpha$. 
In  this respect, Solution II$^-$ is thus similar to Solution I: in both cases the length is infinite. 
Evaluating the integral for Solution II$^+$, one finds instead
\eq{
\label{LyIIp}
       L_y=\frac{2R}{\sqrt{\pi}} \frac{\Gamma\left(\frac{\alpha^+}{16}\right)}{ \Gamma\left({\frac{\alpha^+}{16}}+{\frac 12}\right)}\approx 4.7\, R\,.
}
We note that in the Einstein frame the situation is qualitatively similar: Solutions I and II$^-$ exhibit an infinite proper length $L_y$, while Solution II$^+$ a finite one ($\approx 3.9 R$).

Since \eqref{LyIIp} is finite, there are two end-of-the-world walls at finite distance from one another (and thus within the same effective description).
In view of the cobordism conjecture and by taking the logarithmic singularity at $|r|=R/2$ into account, it is natural to assume that there will be two corresponding ETW 7-branes located in the $(r,y)$-plane at the two distinguished positions
\eq{
           {\rm ETW}_1:\ (r,y)=(R/2,-\infty)\,,\qquad  {\rm ETW}_2:\ (r,y)=(R/2,+\infty)\,.
}
In the following, we focus on Solution II$^+$. We can compute the string-frame proper length in the $r$-direction (at fixed $y$). It is given by
\eq{
         L_r=  \int_{-R/2}^{R/2}  dr\, e^{B(r)+{\frac 14}\chi(r)}=2 \int_{-R/2}^{0}  dr\,
         \frac{\left(\sin\left[ {\textstyle \frac{\pi}{2R}} \left(r+\frac R2\right)\right]\right)^{\frac{\mu}{8}+\frac{\alpha^+}{32}}}{
         \left(\tan\left[ {\textstyle \frac{\pi}{4R}} \left(r+\frac R2\right)\right]\right)^{\frac{\alpha^+}{8}+{\frac 12}}}\approx 2.1\, R\,
}
and, despite a singularity of the integrand at $r=\pm R/2$, it comes out finite. In the Einstein frame it is finite as well $(\approx 0.9 R)$.
The area of the compact $(r,y)$ space is then simply (in string frame)
\eq{
                  \text{Area}=\int dr dy\, \sqrt{G_{rr}G_{yy}} = L_r\, L_y \approx  9.9\,  R^2\,
}
and once again is a finite quantity.

Actually, the length $L_r$ also depends on the coordinate $y$ due
to the $y$-dependent part of the warp factor and similarly $L_y$
depends on $r$. We discuss briefly these behaviors below.
First, let us consider how the length of the circle in the $r$-direction
gets warped with the $y$-coordinate
\eq{
  R(y)= e^{V(y)+{\frac 14} \psi(y)}\, R =e^{-\alpha^+ U(y)}\, R
   \underset{y\to \pm\infty}{\longrightarrow}  0\,.
}
The behavior in the Einstein frame is qualitatively similar, $R(y) =
e^{-\frac 54\alpha^+ U(y)}R$.
As for $L_y$, close to the singularity at $r=R/2$ it is multiplied with the exponential of
\eq{
  \label{lengthlystring}
  B(r)+{\frac 14} \chi(r)\simeq {\frac 18}\left({\frac{(\alpha^+)^2}{16}}+{\frac{\alpha^+}{2}}-3\right) \log\rho \approx -0.16 \log\rho,
}
indicating that the proper length of the interval diverges at $\rho\to 0$.
Notice that, on the contrary, in the Einstein frame the warp factor becomes
\eq{
  \label{lengthlyeinstein}
  B(r)\simeq {\frac
    18}\left({\frac{(\alpha^+)^2}{16}}+{\frac{\alpha^+}{4}}+1\right)
  \log\rho \approx 0.26 \log\rho,
}
and thus the proper length goes to zero  as $\rho\to 0$.

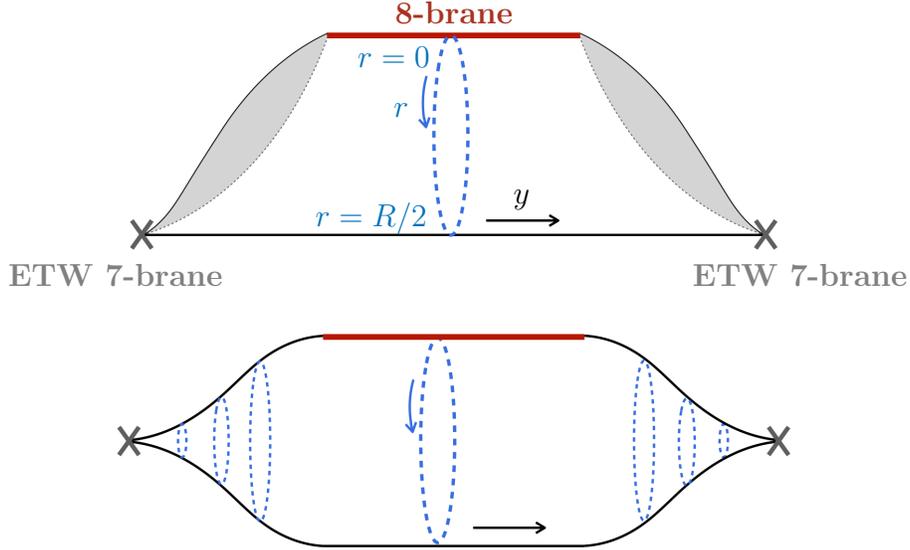
\begin{figure}[ht!]
\centering
  \includegraphics[width=10cm]{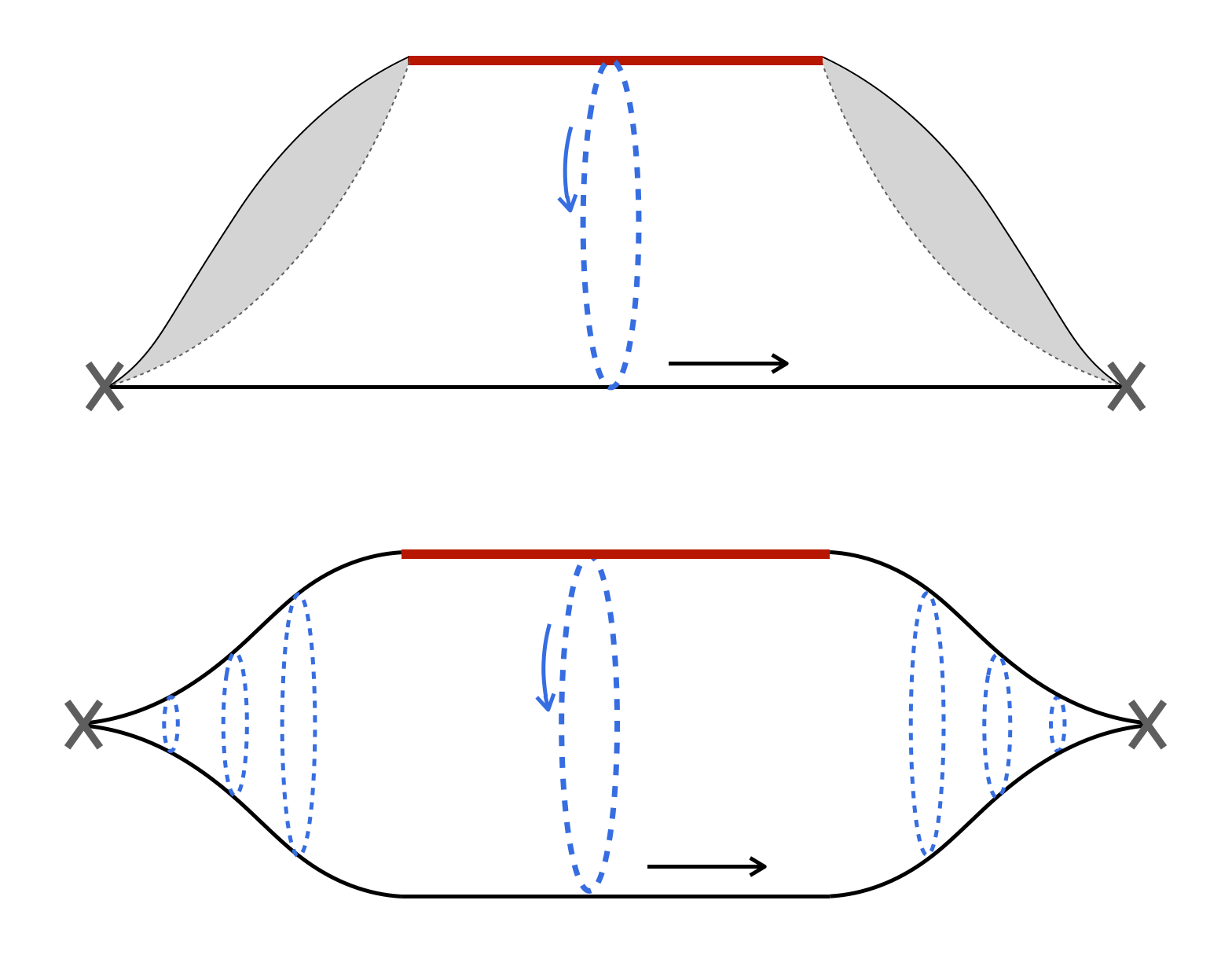}
    \caption{Schematic view of the string frame
      geometry.}
  \label{fig:geom}
  \begin{tikzpicture}[ overlay , remember picture, shorten >=2pt,shorten <=.5pt]
  \node[Mahogany] at (0,9.) {\bf 8-brane};
  \node[RoyalBlue] at (-0.7,7.7) {$r$};
     \node[RoyalBlue] at (-0.8,8.4) {$r=0$};
       \node[RoyalBlue] at (-1.1,6.25) {$r=R/2$};
  \node[black] at (.9,6.5) {$y$};
  \node[gray] at (-4.5,5.5) {\bf ETW 7-brane};
  \node[gray] at (+4.6,5.5) {\bf ETW 7-brane};
  \end{tikzpicture}
\end{figure}

Therefore, at the presumed location of the ETW 7-branes, in string frame the size of the circle in the $r$-direction tends to zero, while the length of the interval in the $y$-direction goes to infinity, such that the area stays finite.
This means that topologically this space is the unreduced suspension of the circle, $S(S^1)=S^2$.
In figure \ref{fig:geom}, we provide a schematic representation of the
solution in the string frame. he two gray circles on the left and on the right hand side of the upper figure actually have zero size and should be considered as points. This is shown explicitly in the lower figure.

\noindent
As we have seen, in the Einstein frame the distance of the two
ETW 7-branes at $r=R/2$ goes to zero, which means that
the space in the $(r,y)$-plane is topologically the
reduced suspension $\Sigma(S^1)=S^2$.

We would like now to relate our setup and the general picture of Local Dynamical Cobordism \cite{Buratti:2021yia,Buratti:2021fiv,Angius:2022aeq}.
Recall that for the distance $\Delta$ to the ETW-defects and their tension $\mathcal{T}$, the following scaling behavior
\eq{
  \Delta^{-n}\sim {\cal T}
}  
was proposed in \cite{Buratti:2021yia}, with $n$ some order one parameter. 
For the present example, one can indeed  express $\Delta$ in terms of the tension ${\cal T}\sim \lambda\exp({\frac 54}\phi_0)$ of the  neutral domain wall
\eq{
       \Delta\sim L_y\sim {\cal T}^{-1},
 }
where we used \eqref{dilatonvev}.
This means that in our case we have the exponent $n=1$.
It was also proposed \cite{Buratti:2021fiv,Angius:2022aeq} that (in units where $\kappa_{10}=1$ and in Einstein frame) the distance $\Delta$  and the scalar curvature ${\mathcal{R}}$ scale with the distance $D$ in field space as
\begin{equation}
  \label{scalingetw}
  \Delta\sim e^{-{\frac{\delta}{2}} D}\,,\qquad |{\mathcal{R}}|\sim e^{\delta D}\
\end{equation}
for a suitable parameter $\delta$.
This is reminiscent of the swampland distance conjecture \cite{Ooguri:2006in}, as it means that close to the end-of-the-world the field goes to infinity with a logarithmic behavior. 

We will check that a relation of the type \eqref{scalingetw} holds in
our setup, both in string and Einstein frame, for a certain value of
$\delta$ different in the two frames.
Consider the ETW 7-brane located at $y=+\infty$ and $|r|=R/2$. 
Since the proper length of the circle in the $r$-direction goes to zero at $y=+\infty$, all trajectories specified by a fixed $r$ and $y\to+\infty$ do reach  the  ETW 7-brane location.
Note that e.g.~trajectories with a fixed finite value of $y$ and $r\to R/2$ do not reach  the  ETW 7-brane location. 
Therefore, for the string-frame distance close to the location of the ETW brane, we can approximate 
\eq{
         \Delta\sim \int_{y}^\infty  dy'\, e^{-{\frac{\pi\alpha}{16 R}} y'}
           \sim e^{-{\frac{\pi\alpha}{16 R}} y}\,.
}
Expressing this in terms of the field distance of $\psi(y)\sim
{\frac{\pi\alpha}{ 16 R}} y$ and its canonical normalized field
$D=\psi(y)/\sqrt{2}$, one finds\footnote{Actually, we are neglecting
  the $r$-dependence in $\Delta$. By including it, we get a power law
  correction of the type (in string frame) $\Delta\sim
  \rho^{(1-2\sqrt{2})}e^{-\sqrt{2} D}$ yielding $\delta=2\sqrt{2}$.}
\eq{
         \Delta\sim e^{-\sqrt{2} D}\,.
}
A similar calculation in the Einstein frame gives $\Delta \sim e^{-\frac{5\pi \alpha}{64}y} \sim e^{-\frac 54 \sqrt 2 D}$ and thus $\delta=\frac 52 \sqrt 2$.
It remains to check whether the second relation in \eqref{scalingetw} is satisfied for these values of $\delta$. 
For the general warped ansatz of the 10D metric \eqref{metricansatz},  the
Ricci scalar is
\eq{
       {\mathcal{R}}=-e^{-2{\cal B}}\Big( 16\, \Box {\cal A} +72\, (\nabla{\cal
           A})^2+2\, \Box{\cal B}\Big)
}        
giving in the Einstein frame $|\mathcal{R}| \sim e^{-2 V(y)} \sim e^{\frac 52 \sqrt 2 D}$, while in the string frame
\eq{
        |{\mathcal{R}}|\sim e^{-2 \left(V(y)+{\frac 14} \psi(y)\right)}\sim  e^{2\sqrt{2}
        D}\,.
}      
The relations \eqref{scalingetw} are thus satisfied in our setup.

\subsection{Cobordism interpretation}
\label{sec_cobord}

Comparing the three solutions, we observe that only Solution II$^+$ features an interval of finite proper length. 
As we will discuss in the next section, this behavior matches precisely with the presence of a 7-brane solution curing the logarithmic singularity and capping off the $y$-direction. 
Such an ETW-brane configuration can be actually predicted by the requirement of a vanishing cobordism group proposed in \cite{McNamara:2019rup}, as we are going to argue below.

We recall that a trivial cobordism group, $\Omega_k=0$, corresponds to the statement that any compact $k$-dimensional quantum gravity background is the boundary of a $(k+1)$-dimensional compact manifold.
This is equivalent to the requirement that (certain) higher-form global symmetries should be absent in the $(d-k)$-dimensional effective theory. 
Thus, the global symmetry must either be gauged or broken.
In the latter case, one needs to introduce a $(d-k-1)$-defect, which is a domain wall in the effective theory.

As was pointed out in this context in \cite{Blumenhagen:2021nmi}, also non-trivial K-theory classes $K^{-k}({\rm pt})\ne 0$ correspond to global symmetries, which are deeply related to cobordism by a generalization of the classic Conner-Floyd theorem \cite{ConnerFloyd, Hopkins1992}. 
In fact, K-theory symmetries are always gauged \cite{Freed:2000ta} and their charges appear in the same charge neutrality condition together with the cobordism invariants, which lie in the image of the Atiyah-Bott-Shapiro orientation. In many cases, this has been identified  with familiar tadpole cancellation conditions in string theory \cite{Blumenhagen:2021nmi}.

In our setup, we always have a spontaneous compactification on a topological $S^1$.
A natural cobordism group to consider would then be $\Omega_1^{\rm Spin}=\mathbb{Z}_2$. 
The choice of Spin structure is dictated by the fact that we want to have fermions in our effective theory. 
Since the group is not trivial, we have a global symmetry, which must be either broken or gauged.
However, only breaking the symmetry would predict a 7-brane defect.

As was argued in \cite{Blumenhagen:2021nmi},  for one of the two string realizations of the 9-dimensional neutral domain wall that we are aware of, namely the type I non-BPS D8-brane, the breaking is not expected. 
In this case, the symmetry has to be gauged. 
Indeed, it is known that type I (non-BPS) Dp-branes are classified by KO-theory \cite{Witten:1998cd, Gukov:1999yn}. 
In particular, a non-BPS D8 brane is associated to KO$^{-1}$(pt)$\,=\!\mathbb{Z}_2$. 
Via the Atiyah-Bott-Shapiro orientation, this group is indeed isomorphic to $\Omega_1^{\rm Spin}$.
However, as proposed in \cite{Blumenhagen:2021nmi}, in this peculiar case the cobordism invariant contribution to the $\mathbb Z_2$-valued tadpole cancellation condition vanishes (mod 2).  
Thus, gauging the global $\mathbb{Z}_2$ symmetry we started with requires an even number of non-BPS D8 brane charges to cancel the tadpole on the KO-theory side.

Therefore, we are led to the proposal that Solution I or II$^-$ describes the backreacted non-BPS D8-brane, classified by KO$^{-1}$(pt).
These solutions do not have an end-of-the-world brane at finite distance, but still always feature a log-singularity.  
We now interpret this as an inconsistency telling us that one should not consider a single such non-BPS D8-brane on a compact space $S^1$, in agreement with the aforementioned gauging.

For the other string realization of the 9-dimensional neutral domain wall, namely the $\overline{D8}/O8$ stack of a non-supersymmetric type IIA orientifold, a natural interpretation is that instead the cobordism symmetry $\Omega_1^{\rm Spin}=\mathbb{Z}_2$ is
broken. 
Indeed, since we are not in type I, we cannot make use of the aforementioned relation between cobordism and K-theory and there is no obvious motivation why the cobordism charge should be gauged. 
On the other hand, breaking such a symmetry predicts the existence of 7-brane defects which can be nicely identified with the ETW-branes of Local Dynamical Cobordism \cite{Buratti:2021yia,Buratti:2021fiv,Angius:2022aeq}.
In our interpretation, this situation would correspond to Solution II$^+$, where in fact the singularities are at finite distance and spacetime closes off.\footnote{In this picture, it is also tempting to identify the topological $S^1$ as the generator of $\Omega_1^{\rm Spin}$.
As we have seen, the size of this $S^1$ shrinks to zero at the positions of the two ETW-branes and thus the boundary conditions for the fermions should be periodic.}

\section{The ETW 7-brane}
\label{sec:etw7}

The cobordism conjecture suggests that the singularity we have seen for Solution II$^+$ can be cured by introducing an appropriate pair of ETW 7-branes. 
What kind of 7-branes could they be?
To clarify this point, we move a step forward and try to construct a new 7-brane solution of the dilaton-gravity equations of motion. Close to its core, this solution should show the features that we need to close-off the singularity found for Solution II$^+$.

Thus, we are looking for a 7-brane solution to the equations of motion \eqref{Einsteineom} and \eqref{dilatoneom} that preserves 8D Poincar\'e symmetry, that has $\log\rho$ singularities close to its core ($\rho \to 0$) and, as figure \ref{fig:geom} suggests, that is non-isotropic in the two transversal directions. 
Indeed, recall that there is one direction in which the length scale vanishes and another direction in which it goes to infinity.
The tension of the brane and the dependence on the dilaton are not yet known, but we are going to determine them.

\subsection{Solution breaking rotational symmetry}

Guided by the aforementioned constraints, we make a non-isotropic ansatz for the Einstein-frame metric 
\eq{
  \label{metricansatzetw}
  ds^2=e^{2\hat{\cal A}(\rho,\varphi)} ds_8^2 + e^{2\hat{\cal B}(\rho,\varphi)}( d\rho^2
  +\rho^2 d\varphi^2),
}
with a separated dependence of the warp factors and the dilaton on the
radial coordinate $\rho$ and the angular coordinate and $\varphi$, i.e. 
\eq{
       &\hat{\cal A}(\rho,\varphi)=\hat A(\rho) + \hat
       U(\varphi)\,,\quad \hat{\cal B}(\rho,\varphi)=\hat B(\rho) +
       \hat V(\varphi)\,,\\
       &\hat\Phi(\rho,\varphi)=\hat\chi(\rho) + \hat\psi(\varphi)\,.
}
The hat is used here to distinguish the various quantities from the similar ones employed in the 9-dimensional neutral domain wall solution.

The equation of motion are analogous to the ones we have seen in the
previous section. 
The one related to the variation  $\delta g^{\mu\nu}$ is
\eq{
  \label{eommunu2}
  \Big( 7 \hat A'' +7{\frac{\hat A'}{\rho}}&+28 (\hat A')^2 +\hat B''+ {\frac{\hat B'}{\rho}}+{\textstyle {\frac 14}}(\hat\chi')^2\Big)\\[0.1cm]
  +&{\frac{1}{ \rho^2}}\Big( 7 \ddot{\hat U} + 28 (\dot{\hat U})^2+ \ddot{\hat V} + {\textstyle{\frac 14}} (\dot{\hat
  \psi})^2\Big)=-\hat\lambda\, e^{a \hat\Phi} \, {\frac{1}{2\pi \rho}}\delta(\rho)\,,
}
where as mentioned above the tension $\hat\lambda$ and the parameter $a$ (giving the 7-brane dependence on the dilaton) are left undetermined for the moment.
 The prime denotes the derivative with respect to $\rho$ and the dot the derivative with respect to $\varphi$.
 For the two variations $\delta g^{\rho\rho}$ and $\delta g^{\varphi\varphi}$ we obtain
\eq{
  \label{eomyyrr2}
   &\Big( 8{\frac{\hat A'}{\rho}}+28 (\hat A')^2 +8\hat A' \hat B' - {\textstyle{\frac 14}}(\hat\chi')^2\Big) +
 {\frac{1}{\rho^2}} \Big( 8 \ddot{\hat U}+ 36 (\dot{\hat U})^2- 8
 \dot{\hat U}\dot{\hat V} + {\textstyle{\frac 14}}
 (\dot{\hat\psi})^2\Big)=0,\\[0.2cm]
 &\Big( 8 \hat A'' +36 (\hat A')^2 -8\hat A' \hat B' +{\frac 14}(\hat
  \chi')^2\Big)+{\frac{1}{\rho^2}}
  \Big(  28 (\dot{\hat U})^2+ 8 \dot{\hat U}\dot{\hat V} - {\frac 14} (\dot{\hat
    \psi})^2\Big)=0\,,\\
}
and for  the off-diagonal $\delta g^{\rho\varphi}$
\eq{
  \label{eomyr2}
  &8{\frac{\dot{\hat U}}{\rho}} -8 \hat A' \dot{\hat U}  +8 \hat B' \dot{\hat U} +8 \hat A'
  \dot{\hat V} -{\frac 12} \hat\chi' \dot{\hat \psi}=0  \,.
}
The dilaton equation of motion becomes
\eq{
  \label{eomdila2}
     \Big(\hat\chi''+{\frac{\hat \chi'}{\rho}} +8\hat A'\hat
     \chi'\Big)+{\frac{1}{ \rho^2}}\Big(\ddot{\hat \psi} +8 \dot{\hat U} \dot{\hat
       \psi}\Big)={2a}\lambda \, e^{a \hat\Phi} \, {\frac{1}{2\pi \rho}}\delta(\rho)\,.
}
Note that taking the sum of the two equations in \eqref{eomyyrr2} gives the simple
equation
\eq{
  \label{eomsum2}
        8\Big( \hat A'' +{\frac{\hat A'}{ \rho}}+8 (\hat A')^2\Big)+{\frac{8}{ \rho^2}}\Big(\ddot{\hat U} +8
        (\dot{\hat U})^2\Big)= 0\,.
}
All these equations are similar to the ones from the previous section, albeit  featuring  some extra $1/\rho$-terms. For instance, they contain the 2D Laplacian in polar coordinates
\eq{
         \Box\, { F}(\rho,\varphi) = {\frac{1}{\rho}}\partial_\rho
         (\rho \, \partial_\rho F) +{\frac{1}{\rho^2}} \partial_\varphi^2 F\,.
}

We find it quite remarkable that these equations admit a three-parameter bulk solution which is also similar to the one from the previous section. 
Now it is the $\rho$-dependent functions that are of hyperbolic type 
\eq{
  \label{solutionsrhodep}
        \hat A(\rho)&={\frac 18} \log\left( \cosh\Big[ 8\hat{K} \log\Big({\frac{\rho}{\rho_0}}\Big)
        \Big] \right)\,,\\[0.1cm]
        \hat\chi(\rho)&=\hat\alpha \hat A(\rho)
        ,\\[0.1cm]
        \hat B(\rho)&=-\log\Big({\frac{\rho}{\rho_0}}\Big) +\left({\textstyle {\frac{\hat\alpha^2}{32}}-{\frac 72}}\right) \hat A(\rho)\,,
 }
 where at this stage $\hat\alpha$, $\hat{K}$ and the dimensionful parameter $\rho_0$  are  still undetermined.
 Notice the additional $\hat f(\rho)=-\log(\rho/\rho_0)$ term in the solution for the warp
 factor $\hat B$.
The angle dependent solutions are instead
\eq{
\label{solutionsphidep}
        \hat U(\varphi)&={\frac 18} \log\big\vert \cos (8\hat{K}\varphi) \, \big\vert,\\[0.1cm]
        \hat\psi(\varphi)&={\frac{\hat\alpha}{8}} \log\big\vert \cos(8\hat{K}
        \varphi)\,  \big\vert
         \pm 2 \log\Big\vert \tan\left(4\hat{K}\varphi+{\frac{\pi}{4}}\right) \Big\vert, \\[0.1cm]
         \hat V(\varphi)&=-\left({\textstyle {\frac{\hat\alpha^2}{32}}-{\frac{9}{2}}}\right)\hat U(\varphi)
         +{\frac{\hat\alpha}{16}} \hat\psi(\varphi)\,,
}        
where due to the periodicity of the cosine-function we have $\varphi\in [0,\frac{\pi}{4\hat{K}}]$. 
The sign in the second line is in fact a free parameter, so that the solution comes in a pair. 
Flipping such a sign is the same as shifting the argument of the $\tan$-function by $-\pi/2$, which is the same as flipping the sign of $\hat{K}$.
We denote this pair as ETW $7^\pm$-branes and from now on we invoke the freedom to choose $\hat{K}>0$.

Finally, each of the quantities in the solutions \eqref{solutionsrhodep} and \eqref{solutionsphidep} can be shifted by arbitrary integration constants, which we omitted for convenience. We will see below that matching with our desired delta source configuration can be achieved by fixing one of these constants.

\subsection{Brane sources}

Given the bulk solution, we would like now to study the behavior at
the boundary. To this end, we include the $\delta$-function  source on
the right hand side of the equations of motion.

First, we notice that  $\hat f=\log\rho$ is the 2D Green's function satisfying
\eq{
        \Box_\rho \hat f \equiv  \hat f''+ { \frac{\hat f'}{\rho}}= {\frac{1}{\rho}}\delta(\rho)=2\pi \delta^2(\vec y)\,.
}
To understand where potential $\delta$-functions can appear, we do not initially specify $\hat A(\rho)$. Rather, we introduce just the restricted ansatz
\eq{
  \label{ansatzrestr}
        \hat\chi(\rho)&=\hat\alpha \hat A(\rho)\,,\qquad
        \hat B(\rho)=\underbrace{-\log(\rho/\rho_0)}_{\tilde B(\rho)} +\left({\textstyle {\frac{\hat\alpha^2}{32}}-{\frac 72}}\right) \hat A(\rho),
 }
together with the solution \eqref{solutionsphidep} for the angle
dependent functions, and we insert them into the equations of motion to obtain
\eq{
  \label{emowithsource}
 \delta G^{\mu\nu}&:\quad\phantom{ii}  \Box_\rho\tilde B +
 \left({\textstyle {\frac{\hat\alpha^2+112}{32}}}\right)
 \left(\Box_\rho\hat A + 8 (\hat A')^2 -{\frac{8\hat{K}^2}{\rho^2}} \right) =\, ''\delta(\rho)''\,,\\
 \delta G^{\varphi\varphi}&:\quad  8\,\left(\Box_\rho\hat A
+ 8 (\hat A')^2 -{\frac{8\hat{K}^2}{\rho^2}} \right) =\, ''\delta(\rho)''\,,\\
\delta\Phi&:\quad  \hat\alpha\, \left(\Box_\rho\hat A+ 8 (\hat A')^2
  -{\frac{8 \hat{K}^2}{\rho^2}} \right)
=\, ''\delta(\rho)''\,,\\
}  
with the $\delta G^{\rho\rho}$ and  $\delta G^{\rho\varphi}$ gravity
equations of motion leading to no source term. Here, with $''\delta(\rho)''$ we indicate potential source terms with generic coefficients.

Clearly, the term $\Box_\rho\tilde B$ in the first equation  generates the source term in \eqref{eommunu2} for
$a=0$ and $\hat\lambda=2\pi$.
The second potential contribution on the right hand side of \eqref{emowithsource} is related to $\hat A$, which enters
always with the specific combination
\eq{
  \label{AAsource}
        \hat A'' +{\frac{\hat A'}{\rho}} + 8 (\hat A')^2 -{\frac{8\hat{K}^2}{\rho^2}} =0\,, \qquad {\rm for} \ \rho\ne 0\,.
}
Due to  $\hat A(\rho)\simeq -{\hat{K}}\log\rho/\rho_0$ for $\rho\ll \rho_0$ this could lead to another two-dimensional $\delta$-source.        
However, from \eqref{emowithsource} one would deduce that this source must come from  an 8-brane wrapping the $\varphi$ direction. 
Another possibility is that this term is zero even at the core, $\rho=0$, so that the right hand side of the last two equations in \eqref{emowithsource} is everywhere identically vanishing.
This is preferable in our case, as it would avoid the introduction of yet another 8-brane.

To see that this possibility can indeed occur, notice that the first
two terms in \eqref{AAsource} recombine into
  $\Box_\rho \hat A$ and thus definitely give a $\delta^2$-term so that integrating over a disc of small radius $\varepsilon_0$ yields
\eq{
  \label{werder}
          \int_{D_{\varepsilon_0}}  d\rho d\varphi\, \rho\ \Box_\rho \hat A
            =-{2\pi \hat{K}}\,.
}            
Therefore, this needs to be cancelled by the last
two terms in \eqref{AAsource}, in order to avoid the appearance of additional sources.
Note that all four terms scale like $\pm 1/\rho^2$ close to the core.
To obtain the desired cancellation, we recall that we still have the freedom to choose arbitrary integration constants in the ansatz, as mentioned at the end of the last section.
Actually, after a redefinition of the coordinates $x_\mu$ and $\rho$, we are left with one physical integration constant that is usually identified with $\hat\phi_0$.
Hence, we supplement our ansatz for $\hat A(\rho)$ by
\eq{
  \label{Ageneral}
   \hat A(\rho)={\frac 18} \log\left( \cosh\Big[ 8\hat{K} \log\Big({\frac{\rho}{\rho_0}}\Big)
        \Big] \right)+{\frac{\hat\phi_0}{ \hat\alpha}}\,,
        }
where we have added explicitly such an integration constant.        
Then, we can write
\eq{
      8 (\hat A')^2 -{\frac{8\hat{K}^2}{\rho^2}} &=8\left(\hat A' -{\frac{\hat{K}}{\rho}}\right)\left(\hat A' +{\frac{\hat{K}}{\rho}}\right) \\
       &\simeq -{\frac{16\hat{K}}{\rho}}\, \partial_\rho\!
      \left(\hat A(\rho)+\hat{K} \log\left({\frac{\rho}{\rho_0}}\right)\right),
}
so that integrating over a disc of small radius $\varepsilon_0$ and  invoking Stoke's theorem one gets
\eq{
  \label{hertha}
             \int_{D_{\varepsilon_0}}  d\rho d\varphi\, \rho\, \left(8 (\hat A')^2 -{\frac{1}{8\rho^2}}\right) =-\frac{32\pi \hat{K}}{\hat\alpha}
      \hat\phi_0\,.
 }
 
We conclude that the behavior of the peculiar expression \eqref{AAsource} at the core depends on the value of an integration constant. 
Without specifying this parameter, the expression  is ambiguous.
For the specific value $\hat\phi_0=-\hat\alpha/16$, the term \eqref{hertha} cancels against \eqref{werder} and \eqref{AAsource} vanishes everywhere.
In this case, the only physical $\delta$-source term on the left hand side of
\eqref{emowithsource}  arises  from  the contribution  $\tilde B$, which can be
reproduced by a $7$-brane localized at $\rho=0$ with Einstein-frame action
\eq{
         S_{7}=-T_7 \int d^{10} x\, \sqrt{-g}\; \frac{\delta(\rho)}{
         2\pi \rho}\,.
}       
Since there is no contribution in the dilaton equation, we need to set
$a=0$ and choose $\hat\lambda=\kappa_{10}^2 T_7=2\pi$. Note that for a 7-brane  $\hat\lambda$ is dimensionless.

\subsection{Comparison to  Solution II}

Let us compare the behavior of the solution close to the core, $\rho=0$, with the singular geometry \eqref{singula} of Solution II$^+$.
The parameter capturing the local behavior close to the ETW-wall
(string frame)  is the factor $\delta=2\sqrt{2}$ in the scaling relation \eqref{scalingetw}. 
We now require that our ETW 7-brane solution scales in the same way close to its core.

The $\rho$-dependent functions \eqref{solutionsrhodep} close to $\rho=0$ read
\eq{
  \label{singulb}
  \hat A(\rho)&\simeq -{\hat K}\log\rho\,,\quad
    \hat\chi(\rho)\simeq -{\hat K\hat\alpha} \log\rho\,,\\
    \hat B(\rho)&\simeq - {\hat K}\left({\textstyle {\frac{\hat\alpha^2}{32}}-{\frac 72}+{\frac{1}{\hat K}}}\right) \log\rho\,.
}
Then, the string-frame distance to the core behaves as
\eq{
     \Delta
     \sim \int_{0}^\rho  d\rho'\, e^{\hat B(\rho')+{\frac{1}{4}}\hat\chi(\rho')}\sim \rho^{-\hat K \left(
   {\frac{\hat\alpha^2}{32}}+{\frac{\hat\alpha}{4}}-{\frac 72}\right)}
}
and for $\hat K>0$ it comes out finite only if ${\frac{\hat\alpha^2}{32}}+{\frac{\hat\alpha}{4}}-{\frac 72}<0$.
This distance can be expressed in term of the canonically normalized field $D=\hat\chi/\sqrt{2}$ as
\eq{
       \Delta\sim \exp\left(  {\textstyle {\frac{\sqrt 2}{ \hat\alpha}}}  \left(
  {\textstyle {\frac{\hat\alpha^2}{32}}+{\frac{\hat\alpha}{4}}-{\frac 72}}\right)  D    \right),
}
so that, recalling the scaling relation \eqref{scalingetw}, we get $\delta=2\sqrt 2$ (with $\Delta$ finite) precisely for
\eq{
        \hat\alpha=\alpha^+=4(4\sqrt 2 -5)\,.
      }
As for the scalar curvature, we find close to $\rho=0$ and in string frame
\eq{
                     |{\mathcal{R}}|\sim {\frac{1}{\rho^2}} e^{-2\left(\hat B(\rho)+{\frac 14}\hat\chi(\rho)\right)}\sim \rho^{2\hat K \left(
   {\frac{\hat\alpha^2}{32}}+{\frac{\hat\alpha}{4}}-{\frac 72}\right)}\sim
e^{2\sqrt{2} D},
}
which is also consistent with the scaling behavior
\eqref{scalingetw}.\footnote{We also checked that the scaling
relations  \eqref{scalingetw} are obeyed in Einstein-frame. In this case we found  $\delta  = \frac 52 \sqrt 2$, which consistently
leads to $\hat \alpha = \alpha^+$, as well.}
Thus, close to the core, the coordinate $\rho$ in the 7-brane  solution plays the role of the coordinate $y$ (or better $\Delta$) in the original domain wall solution, for $y\to\pm \infty$. Similarly, the $\varphi$-coordinate for the ETW 7-brane solution \eqref{solutionsphidep} is related to the periodic $r$-coordinate of the domain wall \eqref{sol2r} via $8\hat K\varphi=8K(|r|-3R/2)={\pi\over 2R}|r|-{3\pi\over 4}$. Therefore, depending on $\hat K$, the former domain wall coordinate $r$ parametrizes a segment of the $\varphi$ circular coordinate.
For convenience we choose the value $\hat K=1/8$ from now on, so that $\varphi$ has periodicity $2\pi$. 
Hence, we conclude that close to the core the non-isotropic ETW 7-brane solution has the right properties to close-off the singularity found in the  neutral domain wall solution of the previous section.

It is instructive to  compare the set of parameters in of Solution II$^+$ and of the ETW 7-brane solution.
Initially, for the neutral domain wall the bulk solution admitted five parameters,
\eq{
         \alpha,\, K,\, R,\, e^{\phi_0},\, \lambda\,.
}
Implementing the boundary conditions from the domain wall and requiring a finite size of the spontaneously compactified longitudinal direction fixed $\alpha=4(4\sqrt 2-5)$ and led to the two conditions
\eq{
  K={\frac{\pi}{16R}}\,,\qquad
  e^{{\frac 54}\phi_0}\sim {\frac{1}{\lambda R}} \,.
}
Finally,  we were  left with two unconstrained parameters like
e.g.~the radius $R$ and the overall scale of the string
coupling constant $g_s=e^{\phi_0}$.  In a concrete string theory setting, the tension
would  also be fixed, of course. Note that for large radius $R$ the
string coupling becomes small so that one expects to have good control
over the employed low energy effective action, which is just dilaton
gravity in this case.

For the 7-brane, we also started with five parameters in the bulk (even if $\hat \phi_0$ was introduced explicitly afterwards),
\eq{
         \hat\alpha,\, \hat K,\, \rho_0,\, e^{\hat\phi_0},\, \hat\lambda\, ,
}
but the boundary condition fixed the tension $\hat\lambda=2\pi$ and gave rise to one relation
\eq{
  \phi_0\sim \hat\alpha,
}  
so that one is left with three free parameters.
Requiring that the 7-brane closes off the singularities present in the
neutral domain wall solution (i.e.~that it is really the ETW-brane)
fixed the parameter $\hat\alpha=\alpha^+$.
Hence, at this stage we are also left with two free parameters, namely
$\hat K$ and the radial scale $\rho_0$.
In contrast to the domain wall solution II$^+$,
we now face  a  fixed value of the string coupling constant
$g_s=\exp(-\hat\alpha/16)\approx 0.85$, which is at the boundary of
having control over the utilized low energy effective action. However,
the dilaton-gravity solution per se is valid for arbitrary values of $\hat\alpha$ and
consequently $g_s$.
Therefore, we can retain perturbative control and are thus confident
that the solution captures some physical features of the ETW 7-brane.

\subsection{Geometry of the ETW 7-brane}

Let us now discuss the geometry around the ETW $7^\pm$-branes in more detail.
First, we display the $\rho$-dependent functions \eqref{ansatzrestr} with \eqref{Ageneral} in figure \ref{fig:hatAchiB} below.

\vspace{0.1cm}

\begin{figure}[ht]
  \centering
  \includegraphics[width=4.2cm]{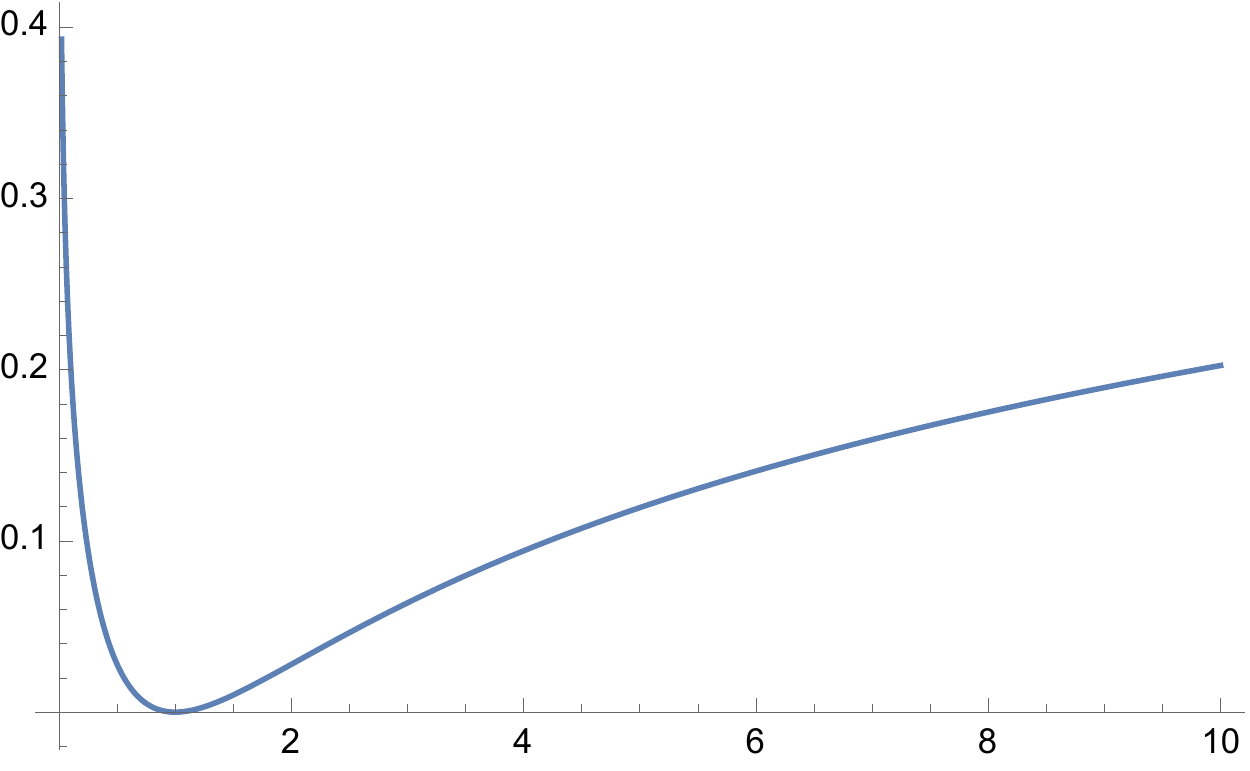}\hspace{0.4cm}
  \includegraphics[width=4.2cm]{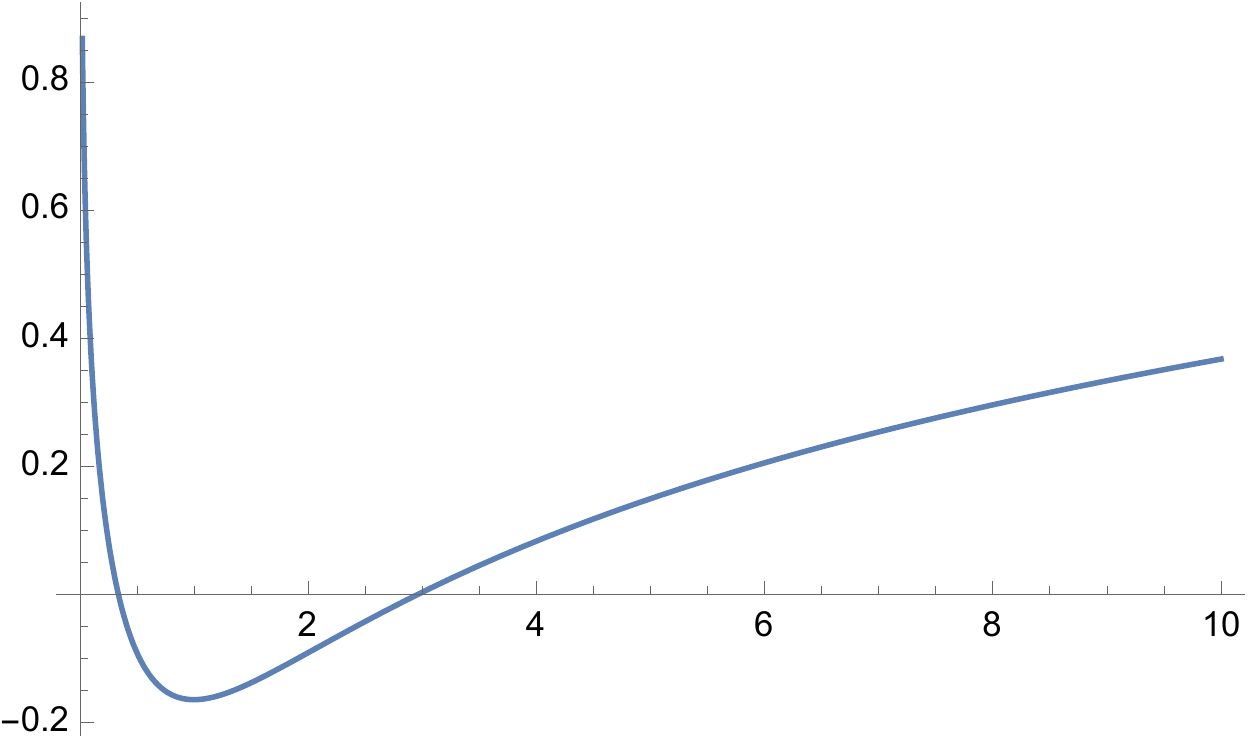}\hspace{0.4cm}
  \includegraphics[width=4.2cm]{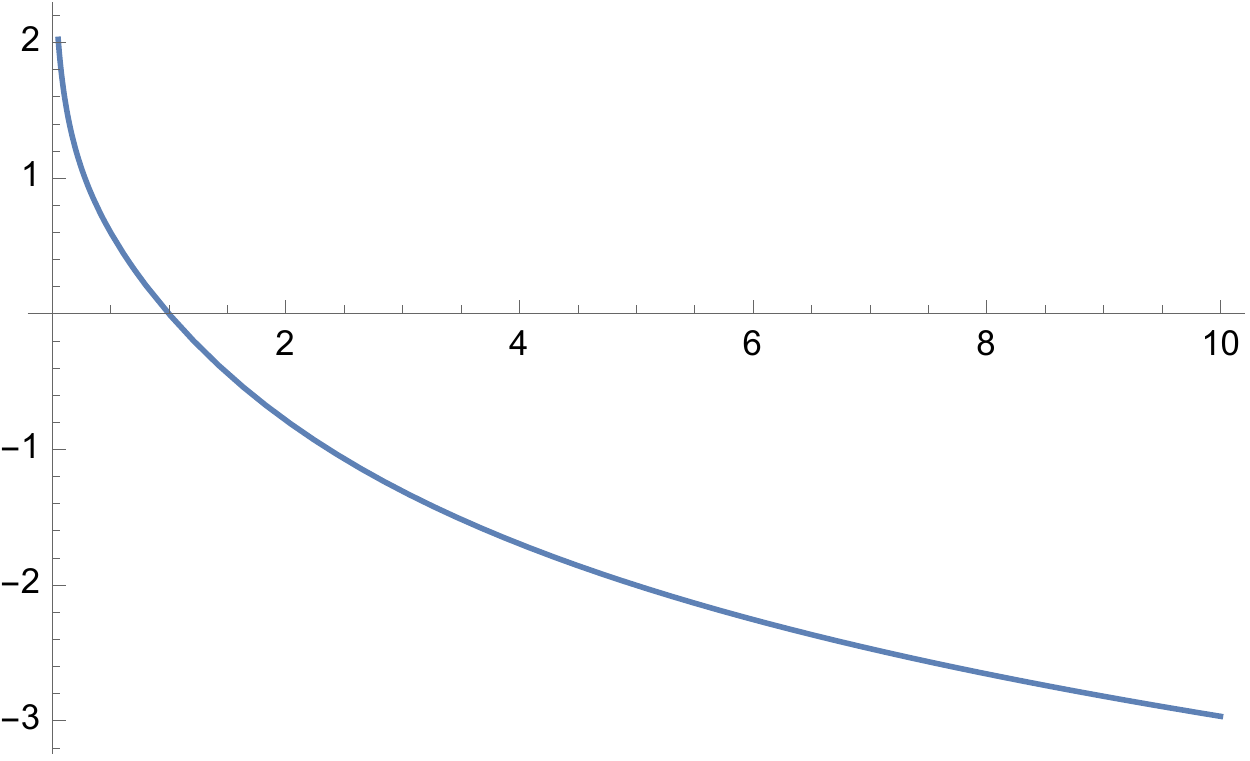}
\begin{picture}(0,0)
    \end{picture}
    \caption{The three functions $\hat{A}(\rho), \hat\chi(\rho)$ and
      $\hat{B}(\rho)$. Here we have chosen $\rho_0=1$ and $\phi_0=-\hat\alpha/16$. }
  \label{fig:hatAchiB}
\end{figure}

\noindent
The figure reflects the just discussed $\log\rho$ behavior close to the core, but except from this no further singularities appear.
Let us also determine the finite proper length of the radial direction
in string frame, which is (we always set $\hat K = \textstyle \frac 18$ below)
\eq{
         L_\rho&=  \int_{0}^\infty  d\rho\, e^{\hat B(\rho)+{\frac 14}\hat\chi(\rho)}\\
           &=\int_{0}^\infty  d\rho\; {\frac{\rho_0}{\rho}}\,
           \Big(\cosh\Big[ \log\Big({\frac{\rho}{\rho_0}}\Big)\Big]\Big)^{{\frac 18}\left({\frac{\hat\alpha^2}{32}}+{\frac{\hat\alpha}{4}}-{\frac 72}\right)}\approx 7.38\,\rho_0\,.
}

\begin{figure}[ht]
  \centering
  \includegraphics[width=4.4cm]{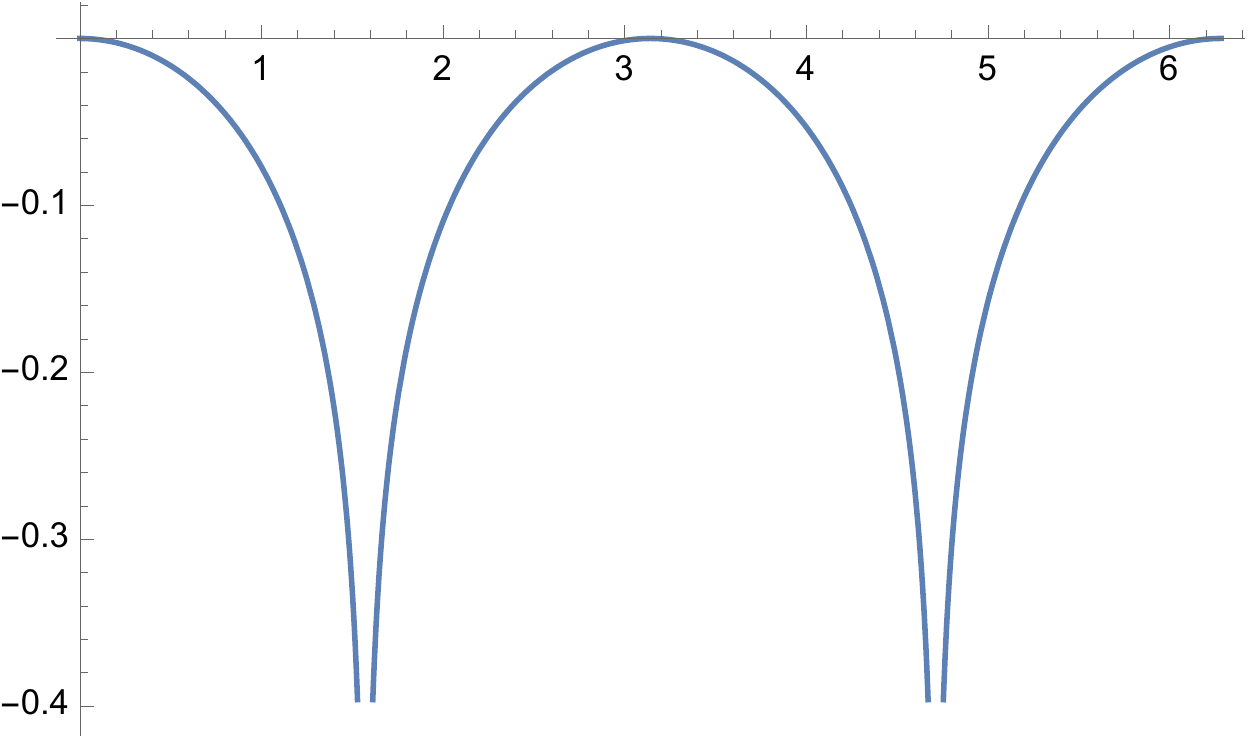}\hspace{0.4cm}
  \includegraphics[width=4.4cm]{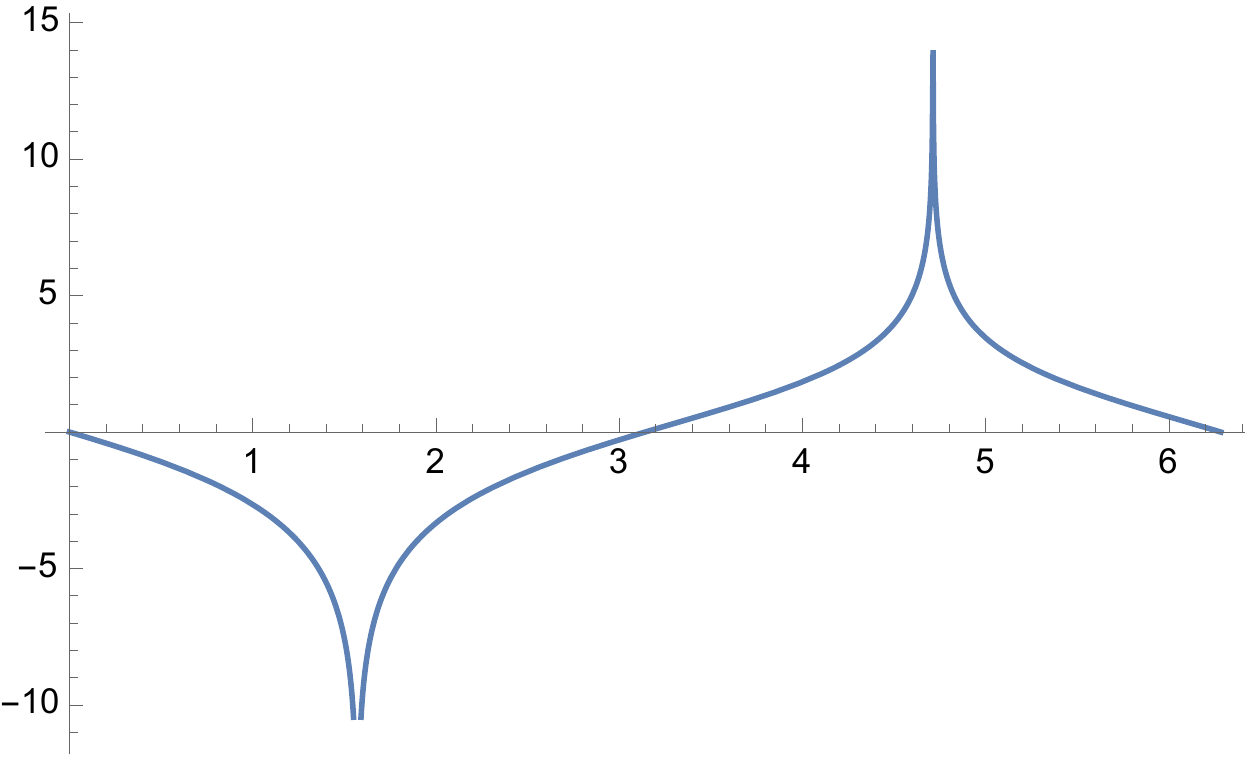}\hspace{0.4cm}
  \includegraphics[width=4.4cm]{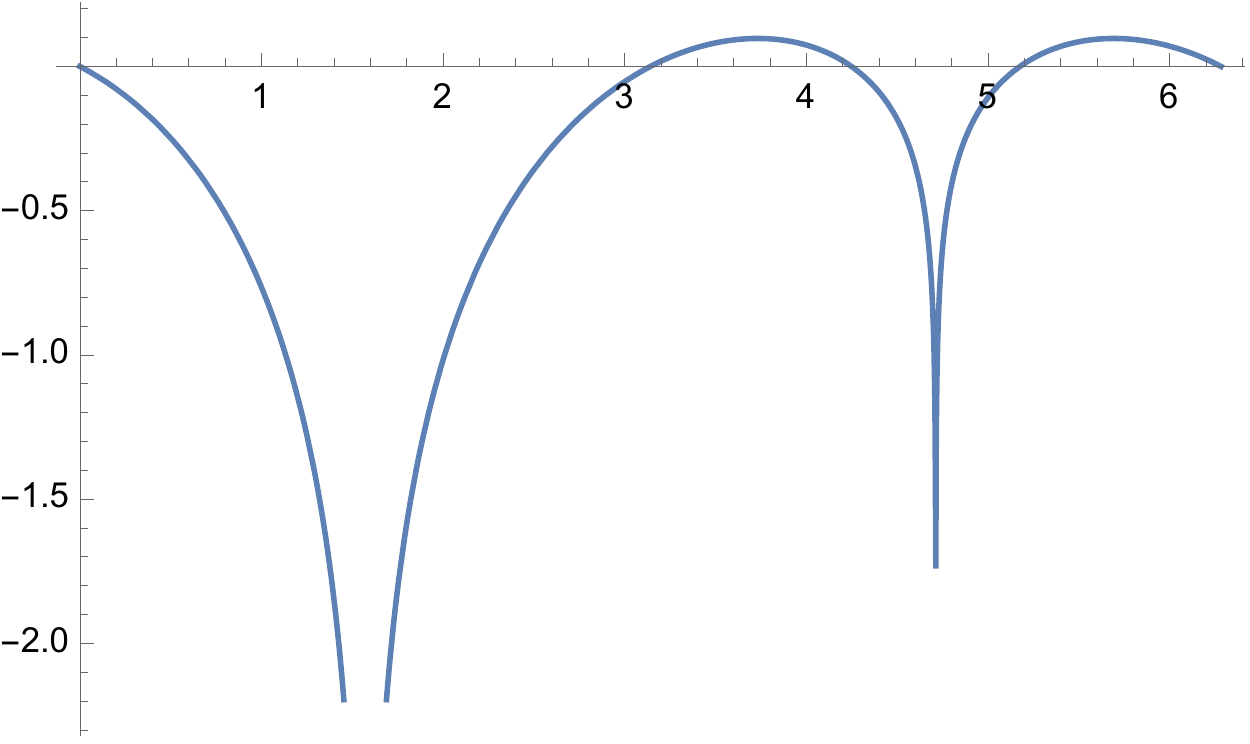}
\begin{picture}(0,0)
    \end{picture}
    \caption{The three functions $\hat{U}(\varphi), \hat\psi(\varphi)$ and
      $\hat{V}(\phi)$ for the ETW $7^-$ brane. For the $7^+$ brane
      the plots are just shifted by $\pi$.}
  \label{fig:hatUpsiV}
\end{figure}

Consider now the more intriguing angular dependence, which we display in figure \ref{fig:hatUpsiV}.
\noindent
First, we see that the solution is non-isotropic, with a singular behavior at the two angles $\varphi_1=\pi/2$ and $\varphi_2=3\pi/2$.
At $\varphi_1$ the string coupling goes to the zero whereas at $\varphi_2$
it diverges.
Next, we compute the proper length of the angular direction in string frame, which is
\eq{
         L_\varphi&=  \int_{0}^{2\pi}  d\varphi\, e^{V(\varphi)+{\frac 14}\psi(\varphi)}\\
          &=\int_{0}^{2\pi}  d\varphi\,
         \Big|\cos \varphi \Big|^{{\frac 18}\left({\frac{\hat\alpha^2}{32}}+{\frac{\hat\alpha}{4}}+{\frac 92}\right)}
         \Big|\tan\left( {\frac{\varphi}{2}}+{\frac{\pi}{4}}\right)\Big|^{\pm {\frac 18}(\hat\alpha+ 4)}
          \approx 1.07\, (2\pi)\,.
}
Thus, both proper lengths $L_\rho$ and $L_\varphi$ are finite but the area, $A= \tilde L_\rho L_\varphi$, is not just the product of the two, as it involves the divergent integral
\eq{
         \tilde L_\rho&=  \int_{0}^\infty  d\rho\, \rho \,e^{\hat B(\rho)+{\frac 14}\hat\chi(\rho)}\to \infty\,.
}
Hence, the ETW 7$^{\pm}$-brane solution is non-compact.

\begin{figure}[ht]
  \centering
   \includegraphics[width=7.0cm]{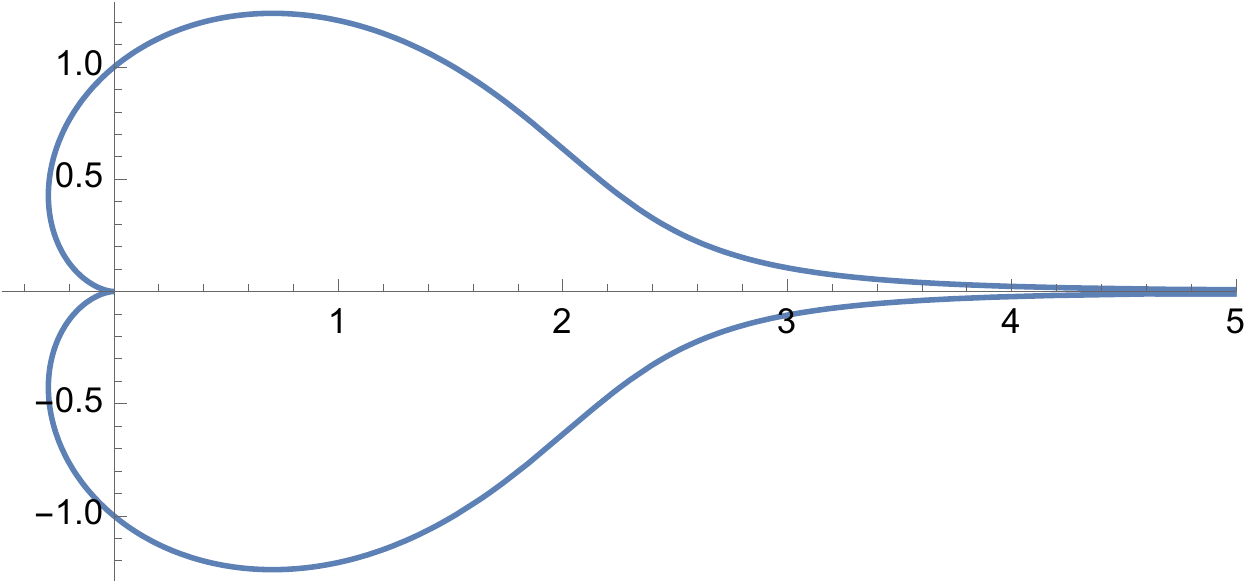}\hspace{0.2cm}
  \includegraphics[width=7.0cm]{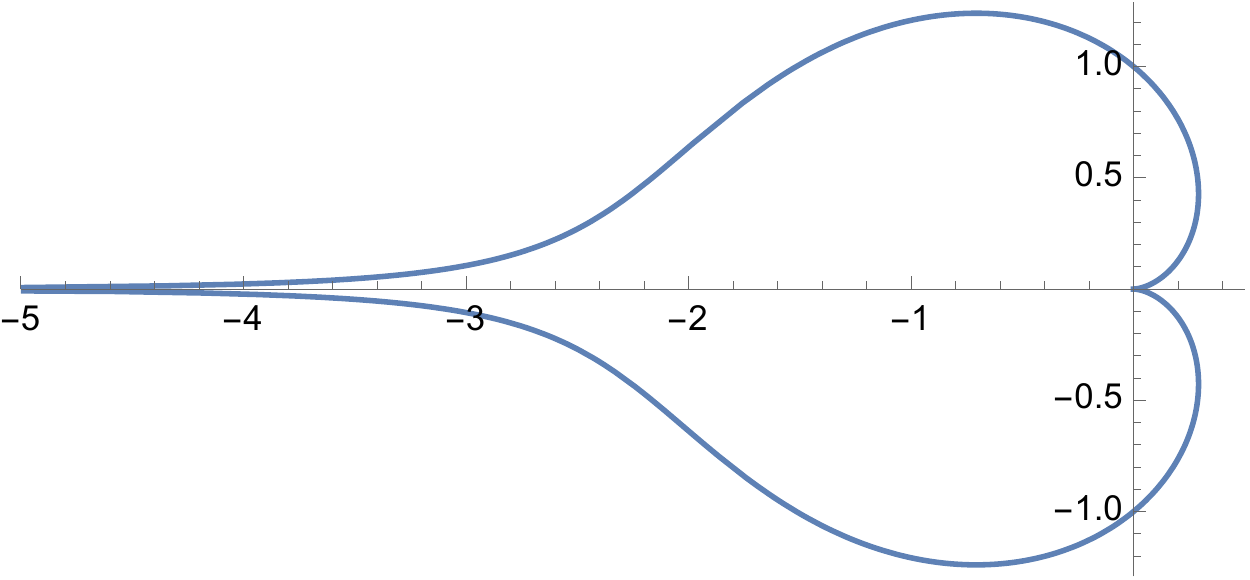}
  \begin{picture}(0,0)
    \put(-150,90){ETW $7^-$ brane}
    \put(-350,90){ETW $7^+$ brane}
    \end{picture}
    \caption{Contour plot of the $\varphi$-dependent warp factor
       $\exp(V(\varphi)+{\frac 14}\psi(\varphi))$ in string frame. For  a more intuitive depiction, we have chosen $\varphi=0$  along the
       negative vertical axis. 
       }
  \label{fig:warpcontour}
\end{figure}
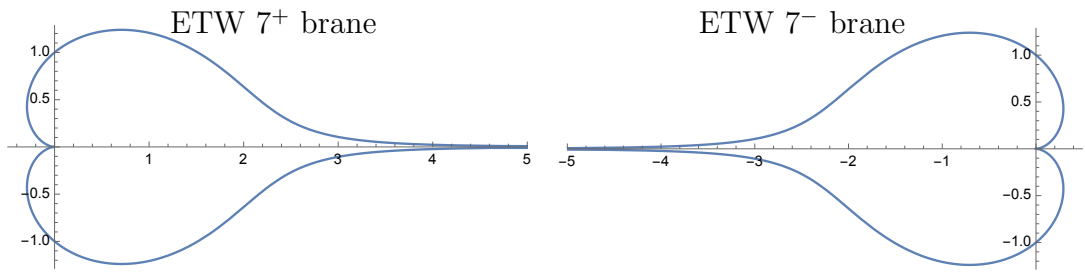

In order to better visualize  what happens to the geometry, in figure \ref{fig:warpcontour} we provide a contour plot of the warp factor 
$\exp(V(\varphi)+{\frac 14}\psi(\varphi))$.
This picture is very suggestive, as one can literally see the structure of the ETW-branes.
There is one weak coupling direction ($\varphi=\mp \pi/2$ for the $7^\pm$ brane), where  the length scale goes to zero and geometry disappears into nothing.
Along the opposite strong coupling direction, the length scale goes to
infinity. This reflects the expected divergent behavior
\eqref{lengthlystring}.\footnote{Doing the same computation in Einstein-frame, as already visible in the rightmost plot in figure \ref{fig:hatUpsiV}, along both opposite directions the radial length scale goes to zero. 
This is consistent with the behavior \eqref{lengthlyeinstein}. }
Note that for each $\varphi\ne \mp\pi/2$ (for the $7^\pm$ brane), the string coupling diverges  as one radially approaches the core at $\rho=0$.

To summarise, we believe we found a very promising candidate for an explicit solution to the leading order string equations of motion describing a defect $7^\pm$-brane that can close-off the singularities from the backreacted non-supersymmetric domain wall solution II$^+$.
As a new characteristic feature, these ETW-branes have a non-isotropic geometry around them, thus breaking the rotational symmetry in the transversal directions familiar  from supersymmetric BPS brane solutions. They have positive tension $\hat\lambda=\kappa^2_{10} T_{7^\pm}= 2\pi$ and in string frame are governed by an action
\eq{
       S=- T_{7^\pm} \int d^{10}x \sqrt{-g}\, e^{-2 \Phi}\,
        \delta^2(\vec r)\,.
 }
To our knowledge, no 7-brane object of this type has so far appeared in the string theory literature.

\section{Conclusions}

In this work, we performed an analysis of previously known classical solutions of dilaton-gravity equations of motion describing neutral domain walls of positive tension, which are necessarily non-supersymmetric. 
They could describe either a non-BPS D8-brane of type I or a $\overline{D8}/O8$ stack of a non-supersymmetric type IIA orientifold.
Among the three solutions, one featured a spontaneous compactification on a finite size interval with two end-of-the-world walls. 
We showed that this Solution II$^+$ fits nicely into the framework of dynamical cobordism, recently proposed in \cite{Buratti:2021yia,Buratti:2021fiv,Angius:2022aeq}.
As a new aspect, we explored a new type of solution of dilaton-gravity equations of motion. Such a solution has the potential to describe the two ETW 7-branes that, according to the cobordism conjecture, should exist in order to close-off the inconsistent background containing just the neutral domain wall.
A novel feature is the non-isotropic nature of this 7-brane solution.

As for future directions, we can point out a couple of open questions, even if some of them might just be artifacts of the effective dilaton-gravity action we are using.
First, what is the role of the other two (also singular) solutions that do not admit ETW walls? 
Since they have a non-compact longitudinal direction and are of topology $S^1\times \mathbb R$, it is conceivable that they are of higher ``energy'' and as such are therefore not stable solutions. Eventually, they might decay to Solution II$^+$.
On the contrary, as we argued in section \ref{sec_cobord}, it could also be that they are on the same level and thus their logarithmic singularity indicates that the global symmetry related to the domain wall charge can only be gauged instead of broken by ETW 7-branes.

Second, we have noticed a compelling mathematical relation between the
spontaneously compactified original bulk solution for the domain wall and the one-dimension-higher non-isotropic ETW 7-brane solution. 
In the first case, the longitudinal Poincar\'e symmetry was spontaneously broken, while in the second case it was the transverse rotational symmetry.
It would be interesting to study whether this is just an accident of this specific example or whether there exists a whole family of such pairs of solutions in e.g. higher codimensions.

Finally, given that our analysis hints at the existence of a possible new (non-supersymmetric) 7-brane in string theory, it would be clearly important to confirm this result with an independent method. Similarly, new string theory defects were predicted in \cite{McNamara:2019rup} and they still call for a better understanding.


\noindent
\paragraph{Acknowledgments:}
We would like to thank R.~Angius, I.~Basile and S.~Raucci for discussions.
The work of N.C.~is supported by the Alexander-von-Humboldt foundation.

\vspace{0.4cm}
\clearpage
\appendix

\section{Additional formal solution}
\label{sec_append}

In this appendix, we collect a new, though arguably not physically interesting, codimension-one solution.
The solution in the bulk is similar to Solution I and given by 
\begin{equation}
    \begin{aligned}
    A(r) &= \frac 18 \log\Big\vert \sin\left[ {\textstyle 8K (|r|
          -{\textstyle {\frac R2}})}\right] \Big\vert,\\
    B(r) &= -\frac{7}{16} \log\Big\vert \sin\left[ {\textstyle 8K (|r|
          -{\textstyle {\frac R2}})}\right] \Big\vert,\\
    \chi(r) &= \phi_0,\\
    U(y) &= \pm K y,\\
    V(y) &= \pm\frac 92 Ky,\\
    \psi(y) &= 0.
    \end{aligned}
\end{equation}
The jump conditions then fix
\begin{equation}
    \cot{(4 K R)} =0, 
\end{equation}
which  admits the minimal solution $K= \pi/(8 R)$. However, this means that the jumps in $A'(r)$ and $B'(r)$ at $r=0$ are vanishing separately so that also the right hand side must be vanishing. 
This implies $\lambda e^{\frac 54 \phi_0} \sim \cot{(4 K R)}=0$ and thus the string coupling vanishes. 
Therefore, this solution does not really describe the backreaction of a positive tension object.

\clearpage

\bibliography{references}  
\bibliographystyle{utphys}


\end{document}